\documentclass{aastex}
\usepackage{amsmath}
\usepackage{amssymb}

\shorttitle{SEP events: trajectory analysis and flux reconstruction with PAMELA}
\shortauthors{Bruno et al.}

\begin{document}
\title{Solar energetic particle events: trajectory analysis and flux reconstruction with PAMELA}

%

\author{
A.~Bruno$^{1,*}$,
O.~Adriani$^{2,3}$,
G.~C.~Barbarino$^{4,5}$,
G.~A.~Bazilevskaya$^{6}$,
R.~Bellotti$^{1,7}$,
M.~Boezio$^{8}$,
E.~A.~Bogomolov$^{9}$,
M.~Bongi$^{2,3}$,
V.~Bonvicini$^{8}$,
S.~Bottai$^{3}$,
U.~Bravar$^{10}$,
F.~Cafagna$^{7}$,
D.~Campana$^{5}$,
R.~Carbone$^{8}$,
P.~Carlson$^{11}$,
M.~Casolino$^{12,13}$,
G.~Castellini$^{14}$,
E.~C.~Christian$^{15}$,
C.~De~Donato$^{12,17}$,
G.~A.~de~Nolfo$^{15}$,
C.~De~Santis$^{12,17}$,
N.~De~Simone$^{12}$,
V.~Di~Felice$^{12,18}$,
V.~Formato$^{8,19}$,
A.~M.~Galper$^{16}$,
A.~V.~Karelin$^{16}$,
S.~V.~Koldashov$^{16}$,
S.~Koldobskiy$^{16}$,
S.~Y.~Krutkov$^{9}$,
A.~N.~Kvashnin$^{6}$,
M.~Lee$^{10}$,
A.~Leonov$^{16}$,
V.~Malakhov$^{16}$,
L.~Marcelli$^{12,17}$,
M.~Martucci$^{17,20}$,
A.~G.~Mayorov$^{16}$,
W.~Menn$^{21}$,
M.~Merg\`e$^{12,17}$,
V.~V.~Mikhailov$^{16}$,
E.~Mocchiutti$^{8}$,
A.~Monaco$^{1,7}$,
N.~Mori$^{2,3}$,
R.~Munini$^{8,19}$,
G.~Osteria$^{5}$,
F.~Palma$^{12,17}$,
B.~Panico$^{5}$,
P.~Papini$^{3}$,
M.~Pearce$^{11}$,
P.~Picozza$^{12,17}$,
M.~Ricci$^{20}$,
S.~B.~Ricciarini$^{3,14}$,
J.~M.~Ryan$^{10}$,
R.~Sarkar$^{22,23}$,
V.~Scotti$^{4,5}$,
M.~Simon$^{21}$,
R.~Sparvoli$^{12,17}$,
P.~Spillantini$^{2,3}$,
S.~Stochaj$^{24}$,
Y.~I.~Stozhkov$^{6}$,
A.~Vacchi$^{8}$,
E.~Vannuccini$^{3}$,
G.~I.~Vasilyev$^{9}$,
S.~A.~Voronov$^{16}$,
Y.~T.~Yurkin$^{16}$,
G.~Zampa$^{8}$,
N.~Zampa$^{8}$,
and V.~G.~Zverev$^{16}$.
}

\affil{$^{1}$ Department of Physics, University of Bari ``Aldo Moro'', I-70126 Bari, Italy.}
\affil{$^{2}$ Department of Physics and Astronomy, University of Florence, I-50019 Sesto Fiorentino, Florence, Italy.}
\affil{$^{3}$ INFN, Sezione di Florence, I-50019 Sesto Fiorentino, Florence, Italy.}
\affil{$^{4}$ Department of Physics, University of Naples ``Federico II'', I-80126 Naples, Italy.}
\affil{$^{5}$ INFN, Sezione di Naples, I-80126 Naples, Italy.}
\affil{$^{6}$ Lebedev Physical Institute, RU-119991 Moscow, Russia.}
\affil{$^{7}$ INFN, Sezione di Bari, I-70126 Bari, Italy.}
\affil{$^{8}$ INFN, Sezione di Trieste, I-34149 Trieste, Italy.}
\affil{$^{9}$ Ioffe Physical Technical Institute, RU-194021 St. Petersburg, Russia.}
\affil{$^{10}$ Space Science Center, University of New Hampshire, Durham, NH, USA.}
\affil{$^{11}$ KTH, Department of Physics, and the Oskar Klein Centre for Cosmoparticle Physics, AlbaNova University Centre, SE-10691 Stockholm, Sweden.}
\affil{$^{12}$ INFN, Sezione di Rome ``Tor Vergata'', I-00133 Rome, Italy.}
\affil{$^{13}$ RIKEN, Advanced Science Institute, Wako-shi, Saitama, Japan.}
\affil{$^{14}$ IFAC, I-50019 Sesto Fiorentino, Florence, Italy.}
\affil{$^{15}$ Heliophysics Division, NASA Goddard Space Flight Ctr, Greenbelt, MD, USA.}
\affil{$^{16}$ National Research Nuclear University MEPhI, RU-115409 Moscow, Russia.}
\affil{$^{17}$ Department of Physics, University of Rome ``Tor Vergata'', I-00133 Rome, Italy.}
\affil{$^{18}$ Agenzia Spaziale Italiana (ASI) Science Data Center, 
I-00133 Rome, Italy.}
\affil{$^{19}$ Department of Physics, University of Trieste, I-34147 Trieste, Italy.}
\affil{$^{20}$ INFN, Laboratori Nazionali di Frascati, 
I-00044 Frascati, Italy.}
\affil{$^{21}$ Department of Physics, Universit\"{a}t Siegen, D-57068 Siegen, Germany.}
\affil{$^{22}$ Indian Centre for Space Physics, 43 Chalantika, 
Kolkata 700084, West Bengal, India.}
\affil{$^{23}$ Previously at INFN, Sezione di Trieste, I-34149 Trieste, Italy. }
\affil{$^{24}$ Electrical and Computer Engineering, New Mexico State University, Las Cruces, NM, USA.}

\altaffiltext{*}{Corresponding author. E-mail address: alessandro.bruno@ba.infn.it.}

\begin{abstract}
The PAMELA satellite experiment is providing first direct measurements of Solar Energetic
Particles (SEPs) with energies from about 80 MeV to several GeV in near-Earth space, bridging
the low energy data by
other 
space-based instruments and the Ground Level Enhancement (GLE) data by the worldwide network of
neutron monitors. Its unique observational capabilities include the possibility of measuring
the flux angular  distribution and thus investigating
possible anisotropies.
This work reports the analysis methods developed to estimate
the 
SEP energy spectra as
a function of
the particle pitch-angle with respect to the Interplanetary Magnetic Field (IMF) direction.
The crucial ingredient is provided by an
accurate simulation of the asymptotic exposition of the PAMELA apparatus, based on a realistic
reconstruction of particle trajectories in the Earth's magnetosphere.
As case study, the results for 
the May 17, 2012 
event are presented.
\end{abstract}


\section{Introduction}\label{Introduction}
SEPs are high energy particles associated with explosive phenomena occurring in the solar atmosphere, such as solar flares and Coronal Mass Ejections. SEP events can significantly perturb the Earth's magnetosphere producing a sudden increase in particle fluxes and, consequently, in the radiation levels experienced by spacecrafts and their possible crew. SEPs constitute a sample of solar material and provide important information about the sources of particle populations, and their angular distribution can be used to investigate the particle transport in the interplanetary medium.

SEP measurements are performed both by in-situ detectors on spacecrafts and by ground-based neutron monitors: while the former are able to measure SEPs with energies below some hundreds of MeV, the latter can only register the highest energy SEPs ($\gtrsim$ 1 GeV) during GLEs. 

New accurate measurements are being 
provided by the PAMELA experiment \citep{SEP2006,MAY17PAPER}. The instrument is able to detect SEPs in a wide energy interval, bridging the energy gap existing between the two aforementioned groups of observations. In addition, PAMELA is sensitive to the particle composition and it is able to reconstruct the flux angular distribution, enabling a clearer and more complete view of the SEP events. This paper reports the analysis methods developed for the estimate of SEP energy spectra as a function of the particle asymptotic direction of arrival. As case study, the results used in the analysis 
of the May 17, 2012 solar event \citep{MAY17PAPER} are discussed.

\section{The PAMELA Experiment}\label{The PAMELA experiment}
PAMELA (a Payload for Antimatter Matter Exploration and Light-nuclei Astrophysics) is a space-borne experiment designed for a precise measurement of the charged cosmic radiation in the kinetic energy range from some tens of MeV up to several hundreds of GeV \citep{Picozza,PHYSICSREPORTS}. The Resurs-DK1 satellite, which hosts the apparatus, was launched into a semi-polar (70 deg inclination) and elliptical (350$\div$610 km altitude) orbit on June 15, 2006; in 2010 it was changed to an approximately circular orbit at an altitude of about 580 km. The spacecraft is 3-axis stabilized; its orientation is calculated by an onboard processor with an accuracy better than 1 deg. Particle directions are measured with a high angular resolution ($<$ 2 deg). Details about apparatus performance, proton selection, detector efficiencies and measurement uncertainties can be found elsewhere (e.g. \citet{SOLARMOD}).

\section{Geomagnetic Field Models}\label{Geomagnetic field models}
The SEP analysis
reported in this work is based
on the IGRF-11 \citep{IGRF11} and the TS07D \citep{TS07D,TS07D2} models for the description of the internal and external geomagnetic field sources, respectively. 
The TS07D is a high resolution dynamical model of the storm-time geomagnetic field in the inner magnetosphere, based on recent satellite measurements.
Consistent with the data-set coverage, the model is valid
up to about 30 Earth's radii (Re). Solar wind and IMF parameters are obtained from the high resolution (5-min) Omniweb database
(http://omniweb.gsfc.nasa.gov/). 

\section{Back-Tracing Analysis}
Cosmic Ray (CR) cutoff rigidities and asymptotic arrival directions (i.e. the directions of approach before encountering the Earth's magnetosphere) are commonly evaluated by simulations, accounting for the effect of the geomagnetic field on the particle transport (see e.g. \citet{SMART_ETAL} and re\-fe\-ren\-ces therein).
Using spacecraft ephemeris data (position, orientation, time), and the particle rigidity ($R$ = momentum/charge) and direction provided by the PAMELA tracking system, trajectories of all detected protons are reconstructed by means of a tracing program based on numerical integration methods \citep{TJPROG,SMART}, and implementing the afore\-mentioned geomagnetic field models.
To reduce the 
computational time, geomagnetically trapped \citep{PAMTRAPPED} and most albedo \citep{ALBEDO} particles
are discarded by selecting
only protons with rigidities $R$ $>$ $R_{min}=10/L^{2}-0.4$ GV, 
where $L$ is the McIlwain's pa\-ra\-me\-ter \citep{McIlwain}.
Each trajectory is back propagated from the measurement location with no constraint limiting the total path-length or tracing time, until: it escapes the model magnetosphere boundaries (Solar or Galactic CRs -- hereafter SCRs and GCRs); or it reaches an altitude\footnote{Such a value refers to the mean production altitude for albedo protons.} of 40 km (re-entrant albedo CRs).
Protons satisfying the latter condition are excluded from the analysis.

The asymptotic arrival directions are evaluated with respect to the IMF direction, with polar angles $\alpha$ and $\beta$ denoting the pitch-angle and the gyro-phase angle, respectively.
To improve the interpretation of results,
the directions of approach and the entry points at the model magnetosphere boundaries can be visualized as a function of the particle rigidity and the spacecraft position. Both Geographic (GEO) and Geocentric Solar Ecliptic (GSE) coordinates are used.

\section{Flux Evaluation}
The factor of proportionality between flux intensities and counting rates, corrected by detector efficiencies, is by definition the apparatus \emph{gathering power} $\Gamma$ (cm$^{2}$sr).
In the case of 
PAMELA, $\Gamma$ is rigidity dependent due to the spectrometer bending effect on particle trajectories\footnote{It decreases with decreasing rigidity $R$ since particles with lower rigidity are more and more deflected by the magnetic field toward the lateral walls of the magnetic cavity, being absorbed before reaching
the lowest plane of the Time of Flight system, which provides the event trigger.}. 
In terms of the zenith $\theta$ and the azimuth $\phi$ angles describing downward-going directions in the PAMELA frame\footnote{The PAMELA reference system has the origin in the center of the spectrometer cavity; the Z axis is directed along the main axis of the apparatus, toward the incoming particles; the Y axis is directed opposite to the main direction of the magnetic field inside the spectrometer; the X axis completes a right-handed system.}:
\begin{equation}\label{gathering_power_formula2}
\Gamma(R)=\int_{-1}^{0} dcos \theta \int_{0}^{2\pi} d\phi \hspace{0.04cm} \left| F(R,\theta,\phi)\hspace{0.04cm}   S(R,\theta,\phi) \hspace{0.04cm}    cos\theta \right|,
\end{equation}
where $F(R,\theta,\phi)$ is the flux angular distribution 
($0\leq F\leq1$), $S(R,\theta, \phi)$ is the apparatus response function in units of area 
and the $\cos\theta$ factor accounts for the trajectory inclination. 

A technically simple but efficient solution for the calculation of the gathering power is pro\-vi\-ded by Monte Carlo methods 
\citep{Sullivan}. 
For isotropic fluxes
$\Gamma$ does not depend on looking direction (i.e. $F$ = 1), and it is usually called the geometrical factor $G_{F}$.
The solid angle is subdivided into a large number of ($\Delta cos\theta,\Delta\phi$) bins, with the angular domain limited to downward-going directions.
For each rigidity, $G_{F}$ can be obtained as:
\begin{equation}\label{differential_GF_tot}
G_{F}(R) \simeq  S_{gen}\hspace{0.04cm} \Delta cos\theta \hspace{0.04cm} \Delta \phi \sum_{cos\theta} \sum_{\phi}\left| \frac{n_{sel}(R,\theta,\phi)}{n_{tot}(R,\theta,\phi)}  \hspace{0.04cm} cos\theta\right|,
\end{equation}
where $n_{tot}$ and $n_{sel}$ are the number of generated and selected trajectories in each ($\Delta cos\theta,\Delta\phi$) bin, and $S_{gen}$ is the area of the used generation surface (see \citet{BRUNO_arXiv} for details). 
An accurate estimate of the PAMELA geometrical factor based on the Monte Carlo approach can be found in \citep{BRUNO_PHD}.

Conversely, in presence of an anisotropic flux exposition ($F\ne const$) the gathering power depends on the flux angular distribution.
Specifically, SCR fluxes can be conveniently expressed in terms of asymptotic 
angles $\alpha$ (pitch-angle) and $\beta$ (gyro-phase angle) with respect to the IMF direction: $F=F(R,\alpha,\beta)$. The corresponding gathering power can be written as:
\begin{equation}\label{gathering_power_formula4}
\Gamma(R)=\int_{0}^{\pi} sin\alpha \hspace{0.04cm} d\alpha  \int_{0}^{2\pi} d\beta \left|   F(R,\alpha,\beta) \hspace{0.04cm} S(R,\theta,\phi) \hspace{0.04cm} cos\theta \right|,
\end{equation}
with $\theta$=$\theta(R,\alpha,\beta)$ and $\phi$=$\phi(R,\alpha,\beta)$. The flux angular distribution $F(R,\alpha,\beta)$ is unknown a priori.
For simplicity, we assume that SCR fluxes depend only on particle rigidity $R$ and asymptotic pitch-angle $\alpha$, estimating the apparatus \emph{effective area} (cm$^{2}$) as:
\begin{equation}\label{gathering_power_formula5}
H(R,\alpha)=\frac{sin\alpha}{2\pi}\int_{0}^{2\pi} d\beta \left| S(R,\theta,\phi)\hspace{0.04cm}   cos\theta \right|,
\end{equation}
by averaging the directional response function over the $\beta$ angle. In case of isotropic fluxes (i.e. independent on $\alpha$) the effective area is related to the geometrical factor by: 
\begin{equation}\label{gathering_power_formula6}
G_{F}(R)=2\pi  \int_{0}^{\pi}d\alpha \hspace{0.04cm}H(R,\alpha).
\end{equation}

$H(R,\alpha)$ can be derived from Equation \ref{differential_GF_tot} by integrating
the directional response function
over the ($cos\theta,\phi$) directions
corresponding
to pitch angles within the interval $\alpha\pm\Delta\alpha/2$:
\begin{equation}\label{MCgfactor_alpha}
2\pi  \int_{\Delta\alpha}d\alpha \hspace{0.04cm}H(R,\alpha) \simeq S_{gen}  \hspace{0.04cm}\Delta cos\theta \hspace{0.04cm} \Delta \phi   \sum_{\theta,\phi\rightarrow\alpha} \left| \frac{n_{sel}(R,\theta,\phi)}{n_{tot}(R,\theta,\phi)} \hspace{0.04cm}cos\theta \right|.
\end{equation}

The used approach is analogous to the one developed for the measurement of geomagnetically
trapped protons \citep{PAMTRAPPED},
but in this case the transformation between local ($\theta$,$\phi$) and magnetic ($\alpha$,$\beta$) angles can not be obtained by simple rotation matrices since it depends on particle propagation in the geomagnetic field;
thus, trajectory tracing methods are necessary.
To assure a high resolution,
$\sim$2800 tra\-je\-cto\-ries (uniformly distributed inside PAMELA field of view - FoV) are reconstructed in the magnetosphere for 1-sec time steps along the satellite orbit and
22 ri\-gi\-di\-ty values between 0.39$\div$4.09 GV,
for a total of about $8\cdot10^{7}$ trajectories for each polar pass ($\sim$23 min).
At a later stage, results are extended over the full
FoV
through a bilinear interpolation.
Since the PAMELA semi-aperture is $\sim$20 deg, the observable 
pitch-angle
range is relatively small (a few deg) except in regions close to the geomagnetic cutoff, 
where 
trajectories
become chaotic and cor\-res\-pon\-ding
asymptotic directions rapidly change with particle rigidity and looking direction; this ends up increasing measurement uncertainties.
Consequently, these zones are excluded from the analysis.

\begin{figure}[!t]
\centering
\includegraphics[width=2.95in,height=3.15in]{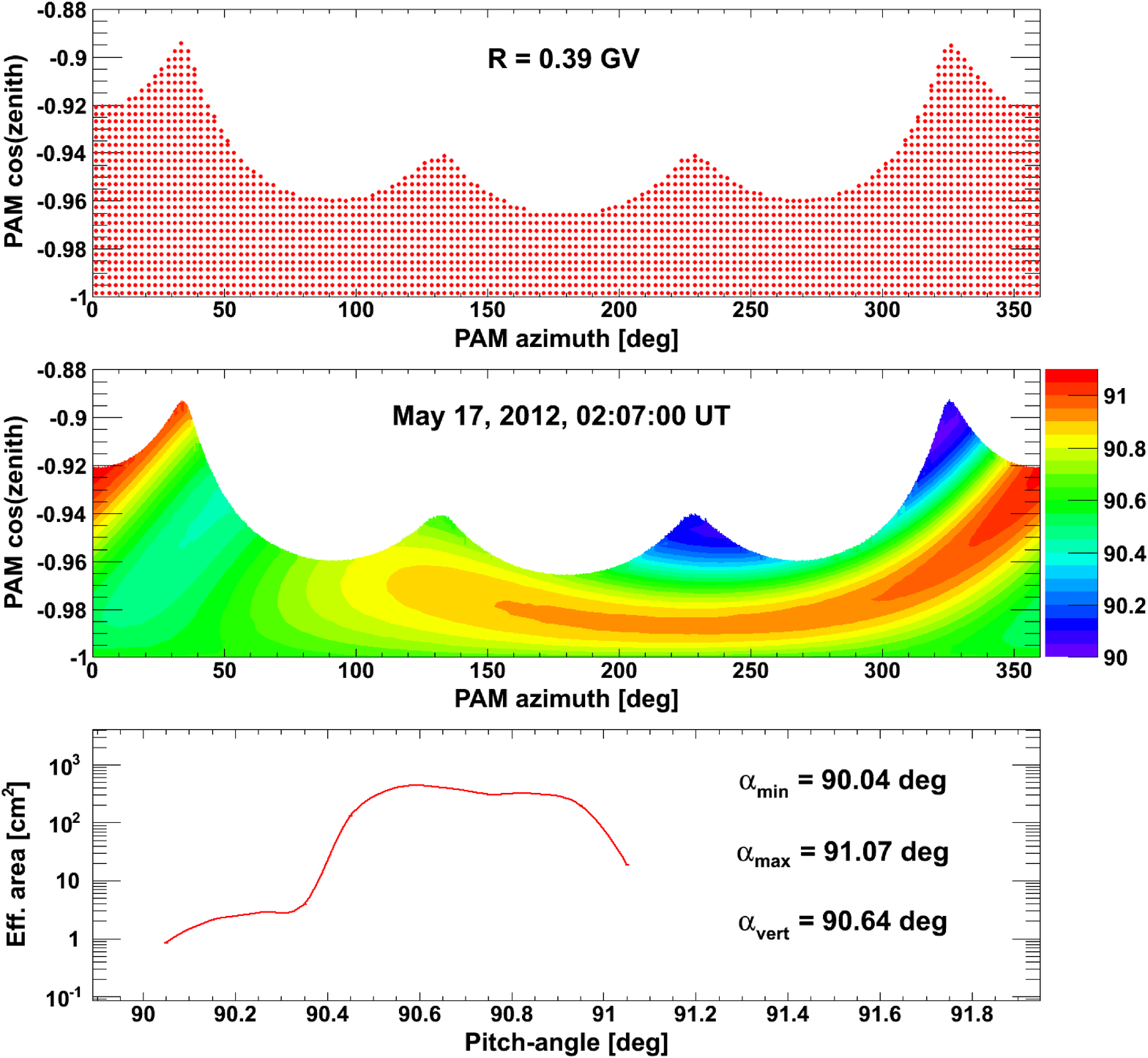}
\includegraphics[width=2.95in,height=3.15in]{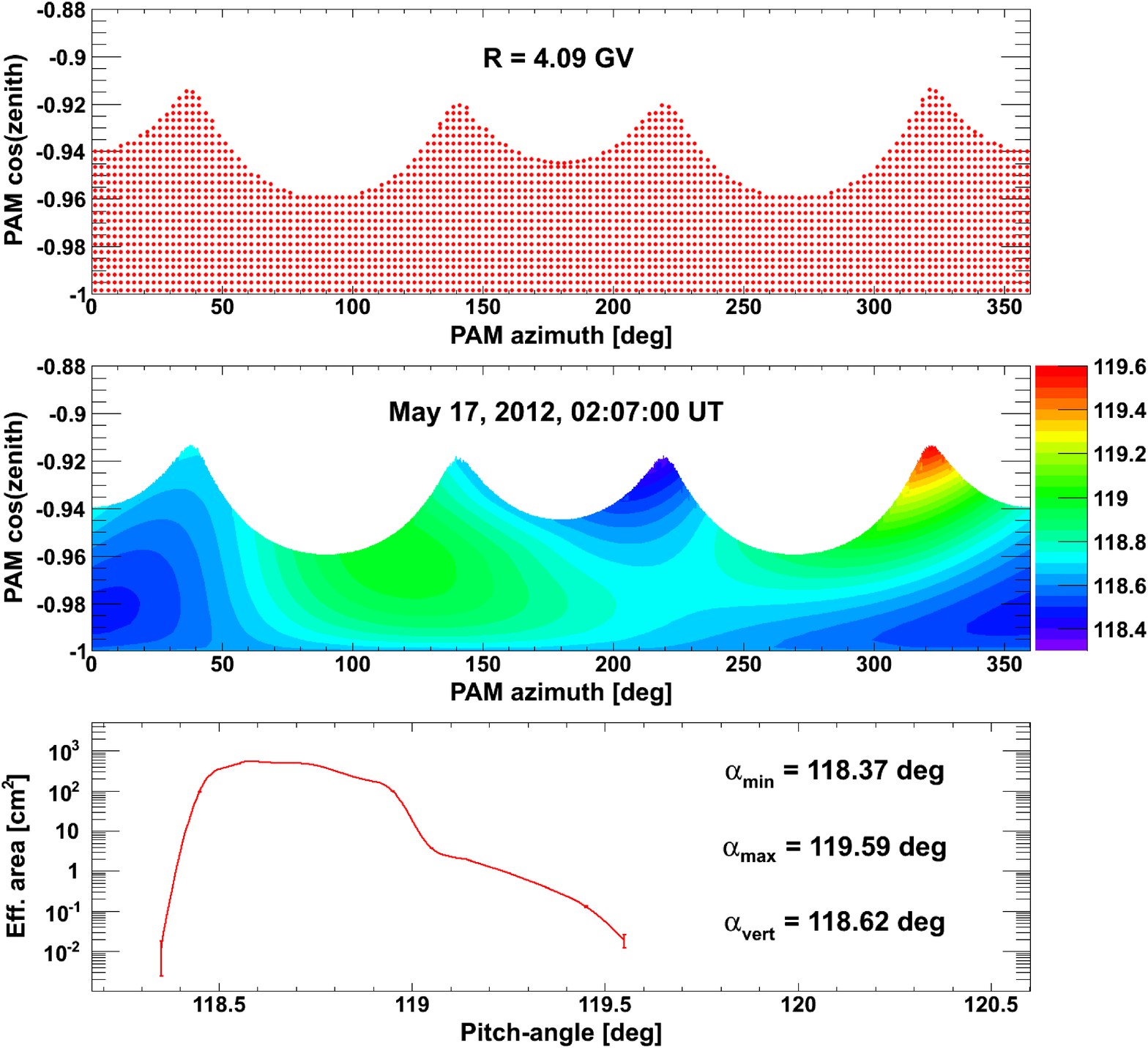}
\caption{Top: distribution of reconstructed directions (red points) inside the PAMELA field of view. Middle: calculated pitch-angle co\-ve\-rage (color code, deg). Bottom: the apparatus effective area as function of the asymptotic pitch-angle; minimum and maximum observable pitch-angles are reported, along with the value corresponding to the vertical direction. Results correspond to 0.39 GV (left) and 4.09 GV (right) protons, 
for a sample orbital position (May 17, 2012, 02:07 UT). See the text for details.}
\label{Figure1}
\end{figure}

\begin{figure}[!t]
\centering
\includegraphics[width=4.5in]{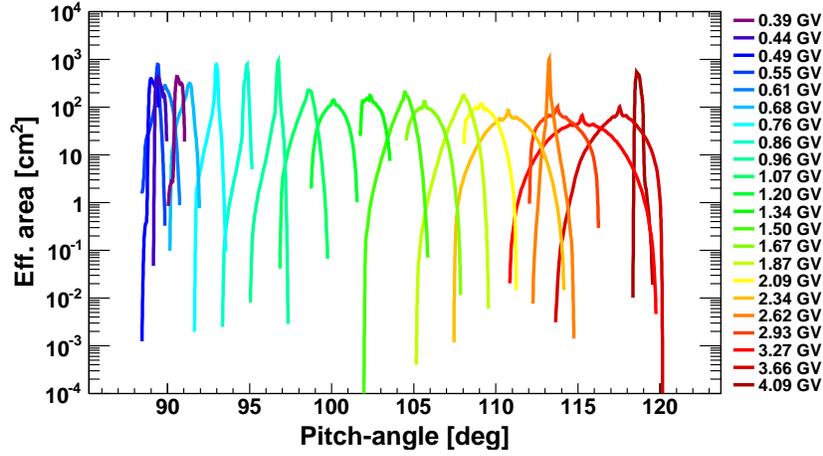}
\caption{The PAMELA effective area as function of pitch-angle at a sample orbital position (May 17, 2012, 02:07 UT),
for different values of particle rigidity (color code).}
\label{Figure2}
\end{figure}

The procedure is demonstrated in Figure \ref{Figure1} for 0.39 GV and 4.09 GV protons (left and right pa\-nels, respectively), at a sample orbital position (May 17, 2012, 02:07 UT).
Top panels report the distributions of reconstructed directions within PAMELA 
FoV\footnote{The covered angular region depends on rigidity as a consequence of the bending effect of the spectrometer; the four peaks reflect the rectangular section of the apparatus.},
with each point as\-so\-cia\-ted to a given asymptotic direction ($\alpha$,$\beta$);
middle panels show the calculated (after interpolation) pitch-angle coverage;
bottom panels illustrate the estimated effective area as a function of 
the explored pitch-angle range.
Effective area
results for 22 ri\-gi\-di\-ty values between 0.39$\div$4.09 GV (color code) are displayed in Figure \ref{Figure2}: the peaks in the distributions
correspond to vertically incident protons.

\begin{figure}[!t]
\centering
\includegraphics[width=4.15in]{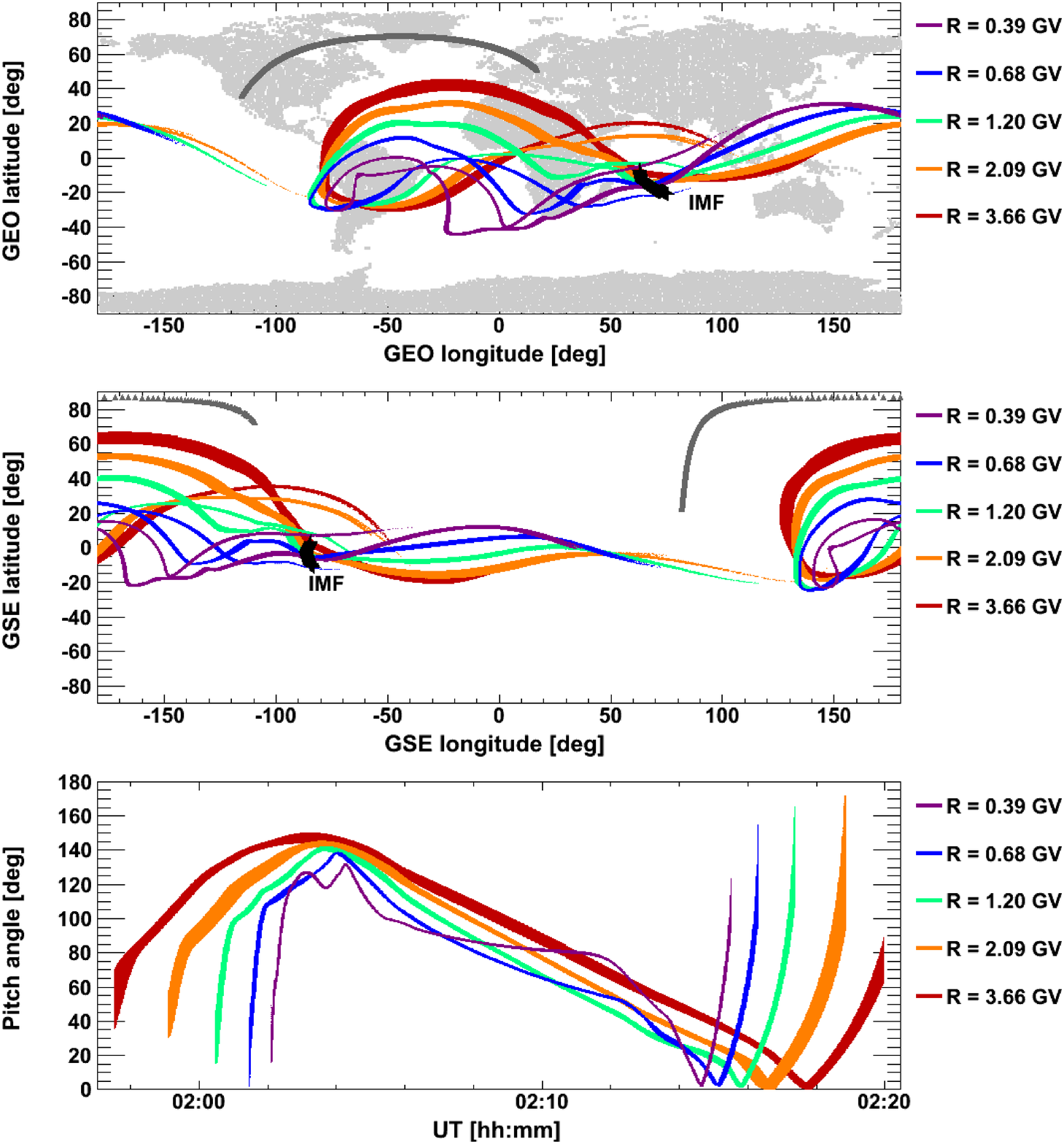}
\caption{Asymptotic cones of acceptance of the PAMELA apparatus for sample rigidity values (color code), evaluated in GEO (top) and GSE (middle) coordinates, and as a function of UT and pitch-angle (bottom). Grey points denote the spacecraft position, while crosses indicate the IMF direction. Calculations refer to the first PAMELA polar pass (01:57$\div$02:20 UT) during the May 17, 2012 SEP event.}
\label{acce_cones2012}
\end{figure}

Figure \ref{acce_cones2012} reports the asymptotic cones of acceptance 
evaluated for the first PAMELA polar pass (01:57$\div$02:20 UT) during the May 17, 2012 SEP event \citep{MAY17PAPER}.
Results for sample rigidity values 
are shown as a function of GEO (top panel) and GSE (middle panel) coordinates;
grey points denote the spacecraft position (northern hemisphere), while crosses indicate the IMF direction.
Finally, the pitch-angle coverage as a function of the orbital position is displayed in the bottom panel.
During the satellite polar pass
the asymptotic cones move in a clockwise direction and
a large pitch-angle interval is covered, approximately ranging from 0 to 145 deg.
In particular, PAMELA is looking at the IMF direction between 02:14 and 02:18 UT, depending on the proton rigidity.

Differential directional flux intensities 
are obtained
at each orbital position $t$ as:
\begin{equation}\label{eq_flux}
\Phi(R,\alpha,t) = \frac{N_{tot}(R,\alpha,t)}{ 2\pi\int\limits_{\Delta R} dR \int\limits_{\Delta \alpha} d\alpha \int\limits_{\Delta t}^{} dt H(R,\alpha,t) },
\end{equation}
where
$N_{tot}(R,\alpha,t)$ is the number of proton counts in the bin $(R,\alpha,t)$, corrected by the detector efficiencies,  
and 
the denominator represents the asymptotic exposition of the apparatus integrated over the selected rigidity bin $\Delta R$.
Averaged fluxes over the polar pass $T=\sum\Delta t$ are evaluated as:
\begin{equation}\label{eq_flux_mean}
\Phi(R,\alpha) =  \frac{N_{tot}(R,\alpha)}{ 2\pi\int\limits_{\Delta R} dR \int\limits_{\Delta \alpha} d\alpha \int\limits_{T}^{} dt H(R,\alpha,t) },
\end{equation}
where $N_{tot}(R,\alpha)=\sum_{T}^{} N_{tot}(R,\alpha,t)$ and the 
exposition is derived by weighting each effective area contribution by the corresponding lifetime 
spent by PAMELA at the same orbital position. 

Final SCR fluxes are obtained by subtracting the GCR contribution from the total measured fluxes.
The GCR component is evaluated by averaging proton fluxes during two days prior to the arrival of SEPs.
We found that GCR
intensities are approximately
isotropic. 
Consequently, the same flux $\Phi_{GCR}(R)$ is subtracted for all pitch-angle bins. 
Statistical errors are obtained by evaluating 68.27\% C.L. intervals for a poissonian signal $N_{tot}(R,\alpha)$ in presence of a background $N_{GCR}(R,\alpha)$.  
Systematic uncertainties related to the reconstruction of asymptotic directions are estimated by introducing a bias in the direction measurement from the tracking system, according to a gaussian distribution with a variance equal to the experimental angular resolution.

\section{Summary and Conclusions}\label{Conclusions}
This paper reports the analysis methods developed for the estimate of SEP energy spectra as a function of the particle asymptotic direction of arrival.
The exposition of the PAMELA apparatus is evaluated 
through
accurate back-tracing simulations based on a realistic description of the Earth's magnetosphere.
As case study, the results of the calculation for the May 17, 2012 event are discussed.
The developed trajectory analysis enables the investigation of flux anisotropies, pro\-viding fundamental information for the characterization of SEPs.
It will prove to be a vital in\-gre\-dient for the interpretation of the solar events observed by PAMELA during solar cycles 23 and 24.

\end{document}